\documentclass{JHEP3}
\usepackage{graphicx}
\def\lesssim{\mathrel{\mathpalette\vereq<}}

\makeatletter
\def\vereq#1#2{\lower3pt\vbox{\baselineskip1.5pt \lineskip1.5pt
\ialign{$\m@th#1\hfill##\hfil$\crcr#2\crcr\sim\crcr}}}
\makeatother
\newcounter{axn}

\newcommand{\bequ}{\begin{equation}}
\newcommand{\eequ}{\end{equation}}
\newcommand{\beqn}{\begin{eqnarray}}
\newcommand{\eeqn}{\end{eqnarray}}
\newcommand{\bit}{\begin{itemize}}
\newcommand{\eit}{\end{itemize}}
\newcommand{\bctr}{\begin{center}}
\newcommand{\ectr}{\end{center}}
\newcommand{\Ls}{\left(}
\newcommand{\Rs}{\right)}

\newcommand{\Ll}{\left[}
\newcommand{\Rl}{\right]}
\newcommand{\LL}{\left.}
\newcommand{\RR}{\right.}
\newcommand{\hsp}[1]{\hspace {#1cm}}

\newcommand{\half}{{1\over2}}
\newcommand{\II}{I$\!$I}

\newcommand{\eg}{{\it e.g.}}
\newcommand{\nn}{\nonumber}
\newcommand{\VEV}[1]{\left\langle #1 \right\rangle}
\def\PR#1#2#3{Phys. Rev. {\bf #1} (#3) #2 }
\def\PRL#1#2#3{Phys. Rev. Lett. {\bf #1} (#3) #2 }
\def\PL#1#2#3{Phys. Lett. {\bf #1} (#3) #2 }

\def\NP#1#2#3{Nucl. Phys. {\bf #1} (#3) #2 }

\def\PTP#1#2#3{Prog. Theor. Phys. {\bf #1} (#3)#2 }

\def\diag{{\rm diag}} 
\newcommand{\eff}{{\rm {eff}}}
\newcommand{\EM}{{\rm {EM}}}
\title{Dynamical symmetry breaking in Gauge-Higgs unification
 of 5D ${\mathcal N}=1$ SUSY theory}

\author{
 Naoyuki Haba\\
 Institute of Theoretical Physics, 
 University of Tokushima, Tokushima 770-8502, Japan \\
}
\author{
 Toshifumi Yamashita \\
 Department of Physics, Kyoto University,
 Kyoto, 606-8502, Japan \\
}

\abstract{
We study the dynamical symmetry breaking in the 
 gauge-Higgs unification of the 5D ${\mathcal N}=1$ SUSY theory,  
 compactified on an orbifold, $S^1/Z_2$. 
This theory identifies Wilson line degrees of freedoms 
 as ``Higgs doublets''. 
We consider $SU(3)_c \times SU(3)_W$ and 
 $SU(6)$ models, in which 
 the gauge symmetries are reduced 
 to $SU(3)_c \times SU(2)_L \times U(1)_Y$ and 
 $SU(3)_c \times SU(2)_L \times U(1)_Y \times U(1)$, 
 respectively, 
 through 
 the 
 orbifolding boundary conditions. 
Quarks and leptons are bulk fields, 
 so that Yukawa interactions 
 can be derived from 
 the 5D gauge interactions. 
We estimate the one loop effective potential 
 of ``Higgs doublets'', and analyze the 
 vacuum structures in these two models. 
We find that the effects of bulk quarks and leptons
 destabilize the suitable electro-weak vacuum. 
We show that the introduction of 
 suitable numbers of 
 extra bulk fields 
 possessing the suitable representations  
 can realize the 
 appropriate electro-weak symmetry breaking. 
}

\preprint{
hep-ph/0402157\\
KUNS-1895 
}

\begin{document}

\section{Introduction}

Recently, there are much progress 
 in the higher dimensional gauge theories. 
One of the most fascinating motivations of 
 considering the higher dimensional gauge theory 
 is that 
 gauge and Higgs fields can be 
 unified\cite{Manton:1979kb,YH,Krasnikov:dt,Lim,Hall:2001zb,Burdman:2002se,HS,CA,mimura1,ghu3,ph-gh}. 
We call this scenario as the 
 gauge-Higgs unification. 
The higher dimensional 
 components of gauge fields
 become scalar fields 
 below the compactification scale. 
These scalar fields are 
 identified with 
 the ``Higgs fields'' 
 in the gauge-Higgs unified 
 models. 
The ``Higgs fields'' have only 
 finite masses of order the 
 compactification scale, 
 since the masses of ``Higgs fields'' are 
 forbidden by 
 the higher dimensional gauge invariance. 
The ``adjoint Higgs fields'' can be 
 induced through the $S^1$
 compactification in 5D theory, 
 while the ``Higgs doublet fields'' 
 can be induced through the orbifold 
 compactifications. 
In order to obtain the 
 ``Higgs doublets'' from the gauge fields 
 in higher dimensions, 
 the gauge group must be lager than the 
 standard model (SM) gauge
 group. 
The gauge symmetries are reduced 
 by the orbifolding boundary conditions of extra 
 dimensions. 
The identification of ``Higgs fields'' as 
 a part of the gauge supermultiplet has 
 been considered in 
 5D ${\mathcal N}=1$ supersymmetric (SUSY) 
 gauge theory 
 whose 5th coordinate is compactified
 on $S^1/Z_2$ 
 orbifold\cite{Lim,Hall:2001zb,Burdman:2002se,HS,mimura1,ph-gh}. 
Also in 6D ${\mathcal N}=2$ SUSY 
 gauge theory\footnote{The 6D ${\mathcal N}=2$ SUSY gauge theory
 corresponds to the 4D ${\mathcal N}=4$ SUSY gauge theory.}, 
 the gauge-Higgs unification 
 has been considered 
 on $T^2/(Z_2 \times Z_2')$ orbifold\cite{Hall:2001zb}.

This paper considers the 
 former scenario, the 5D ${\mathcal N}=1$ SUSY theories 
 compactified on an orbifold, $S^1/Z_2$,  
 in which 
 the Wilson line degrees of freedoms (d.o.f.)
 can be identified as ``Higgs doublets''. 
We consider $SU(3)_c \times SU(3)_W$ and 
 $SU(6)$ models, where 
 the gauge symmetries are reduced 
 to $SU(3)_c \times SU(2)_L \times U(1)_Y$ and 
 $SU(3)_c \times SU(2)_L \times U(1)_Y \times U(1)$,
 respectively, 
 through 
 the orbifolding boundary conditions. 
The case that 
 quarks and leptons 
 are localized on the 4D 
 walls 
 has been studied 
 in Ref.\cite{HHKY}. 
This paper is concentrating on 
 the case that 
 quarks and leptons are 
 bulk fields. 
In this case, 
 Yukawa interactions 
 can be derived from 
 the 5D gauge 
 interactions\cite{Burdman:2002se}. 
We calculate the one loop effective 
 potential\cite{HHKY,HY-proof,HHHK,HHK} 
 of ``Higgs doublets'', 
 and analyze the 
 vacuum structure 
 of the models. 
We find that the effects of bulk quarks and leptons
 destabilize the suitable electro-weak vacuum. 
We show that the introduction of 
 suitable numbers of 
 extra bulk fields 
 possessing the suitable representations  
 makes two appropriate scenarios be possible. 
One is the situation that the 
 one loop effective potential chooses 
 symmetric vacuum at the high energy 
 (compactification) scale, 
 and the electro-weak symmetry breaking 
 is realized by other effects\cite{ph-gh} in 
 the low energy. 
The other is the situation that the 
 one loop effective potential chooses 
  the suitable electro-weak 
 vacuum 
 in a few TeV compactification scale\footnote{
 We should assume the baryon number symmetry 
 to avoid rapid proton decay.
}. 
Here the masses of ``Higgs doublets'' become 
 $\mathcal{O}(100)$ GeV.

\section{Gauge-Higgs unification on $S^1/Z_2$}

At first let us show the notation of 
 the 5D ${\mathcal N}=1$ SUSY gauge theory, 
 which corresponds to the 
 4D ${\mathcal N}=2$ SUSY gauge theory\cite{6DLag}. 
The gauge supermultiplet, $(V,\Sigma)$, of the 
 ${\mathcal N}=2$ SUSY gauge theory 
 is written as 
\begin{eqnarray}
&&V=-\theta \sigma^\mu \bar{\theta}A_\mu + i \bar{\theta}^2 \theta \lambda
    -i \theta^2  \bar{\theta} \bar{\lambda} +
    {1\over2}\bar{\theta}^2\theta^2D ,\\
&&\Sigma={1\over\sqrt{2}}(\sigma+iA_5)+\sqrt{2}\theta \lambda' + 
 \theta^2F.
\end{eqnarray}
$A_5$ is the 5th component of 
 the 5D gauge field. 
In the non-abelian gauge theory,
 the gauge transformation 
 is given by 
$e^V \rightarrow  h^{-1}e^V\bar{h}^{-1}$ and  
$\Sigma \rightarrow h^{-1}(\Sigma + \sqrt{2}\partial_y) h$,
where we denote 
$h\equiv e^{-\Lambda}$, $\bar{h}\equiv e^{-\bar{\Lambda}}$ and  
$V\equiv V^aT^a$, $\Sigma \equiv \Sigma^aT^a$, 
$\Lambda\equiv\Lambda^aT^a$.
Then the action is given by
\begin{eqnarray}
S_{5D}&=&\int d^4x dy \left[{1\over4kg^2}{\rm Tr}\left\{
  \int d^2 \theta W^\alpha W_\alpha
     +h.c.\right\} \right. \nonumber \\ 
&& + \left. \int d^2 \theta {1\over kg^2}{\rm Tr}\left( 
     (\sqrt{2}\bar{\partial}_y  + \bar{\Sigma})e^{-V}
     (-\sqrt{2}\partial_y +{\Sigma})e^V
     +\bar{\partial}e^{-V}\partial e^V
     \right)     \right],
\end{eqnarray}
where ${\rm Tr}(T^aT^b)=k\delta_{ab}$.

As for the hypermultiplet, $(H,H^c)$, 
 they transform $H \rightarrow hH$ and $H^c \rightarrow h^c{}^{-1}H^c$ 
 under the gauge transformation, 
 where $h=e^{-\Lambda^a T^a}$ and
 $h^c=(h^{-1})^T=(e^{\Lambda^a T^a})^T$. 
The action of them is given by
\begin{eqnarray}
S_{5D}^H &=& \int d^4x dy \left[
  \int d^4 \theta (H^c e^V \bar{H}^c + \bar{H}e^{-V}H)+ \right. \nonumber \\ 
&& + \left. \left[\int d^2 \theta 
   \left( H^c \left(m+\left(\partial_y-{1\over\sqrt{2}}\Sigma \right)
  \right)H\right)  +h.c. \right]\right].
\label{5D}
\end{eqnarray}
This means that $\Sigma$ must change the sign 
 under the $Z_2$ projection, $P: y \rightarrow -y$, 
 and bulk constant mass $m$ is forbidden\footnote{
 When the $y$ dependent mass $m(-y)=-m(y)$ is introduced, 
 it makes the zero mode wave-function in the bulk 
 be localized at $y=0$\cite{Kaplan:2001ga}.}.
The field $H^c$ is so-called mirror field, 
 which should have the opposite parity of $P$ compared to $H$, 
 since it is the right-handed field. 
Thus, 
 the parity operator $P$ acts on fields as
\begin{eqnarray}
&& V(y)=P V(-y) P^{-1}, \;\;\;\;\;\; \Sigma(y)= - P \Sigma(-y) P^{-1}, 
\label{5}\\
&& H(y)= \eta T[P] H(-y), \;\;\;\;\;\; H^c(y)= - \eta T[P] H^c(-y). 
\label{6}
\end{eqnarray}
$T[P]$ represents an appropriate 
 representation matrix, for example, 
 when $H$ is the fundamental 
 or adjoint representation, 
 $T[P]H$ means $P H$ or $PHP^\dagger$, respectively. 
The parameter, $\eta$ is like a {\it intrinsic} parity 
 eigenvalue, 
 which takes $\pm 1$. 
As for the reflection around 
 $y=\pi R$ denoted as $P'$, which is reproduced 
 by the product of transformations,
 $y\rightarrow -y$ and $y\rightarrow y+2\pi R$, 
 the bulk 
 fields transform as 
\begin{eqnarray}
&& V(\pi R-y)=P' V(\pi R+y) P'^{-1}, \;\;\;\;\;\; 
   \Sigma(\pi R-y)= P' \Sigma(\pi R+y) P'^{-1}, 
\label{8} \\
&& H(\pi R-y)=\eta'T[P'] H(\pi R+y), \;\;\;\;\;
   H^c(\pi R-y)= \eta' T[P'] H^c(\pi R+y), 
\label{9}
\end{eqnarray}
where $\eta'=\pm$.

We will consider nontrivial 
 $P$ and $P'$ in the gauge group base 
 in order to regard 
 the zero mode components 
 of $\Sigma$ chiral 
 superfield as the ``Higgs doublet''. 
In this case, the $F$-term interaction in Eq.(\ref{5D}) 
\begin{eqnarray}
W_Y \supset H^c \Sigma H,
\label{Yukawa}
\end{eqnarray}
which is invariant under 
 the $Z_2$ projections, 
 can be regarded as the ``Yukawa interactions''. 
The interaction in Eq.(\ref{Yukawa}) 
 connects the chiral and mirror
 fields through the chirality flip, which
 seems to be really Yukawa interaction 
 in the 4D theory.  
This theory proposes that the origin of 
 ``Yukawa interactions''
 exist in the 5D gauge interactions.

\section{$SU(3)_c\times SU(3)_W$ model}

Now let us study the 
 $SU(3)_c \times SU(3)_W$ 
 model, 
 where the Higgs doublets can be identified
 as the zero mode 
 components of 
 $\Sigma$\cite{Lim,Hall:2001zb,Burdman:2002se}. 
We consider the case that  
 quarks and leptons are introduced
 in the bulk to produce 
 Yukawa interactions 
 as Eq.(\ref{Yukawa})\cite{Burdman:2002se,HS,mimura1}. 
We analyze the 
 vacuum structure 
 of this model.

This model takes parities as
\begin{eqnarray}
&&P=P'=\diag(1,-1,-1), 
\end{eqnarray}
in the base of $SU(3)_W$\footnote{We take 
 $P=P'=I$ for $SU(3)_c$.}, 
 which divide $V$ and $\Sigma$ as
\begin{eqnarray}
&&V=\left(
\begin{array}{c|cc}
(+,+)&(-,-)&(-,-) \\ \cline{1-3}
(-,-)&(+,+)&(+,+) \\ 
(-,-)&(+,+)&(+,+) 
\end{array}
\right),\\
&&\Sigma =\left(
\begin{array}{c|cc}
(-,-)&(+,+)&(+,+) \\ \cline{1-3}
(+,+)&(-,-)&(-,-) \\ 
(+,+)&(-,-)&(-,-) 
\end{array}
\right)  .
\end{eqnarray}
This suggests that 
 $SU(3)_W$ is broken down to 
 $SU(2)_L\times U(1)_Y$, 
 and there appear two
 ``Higgs doublet'' superfields 
 as the zero modes of $\Sigma$\footnote{There appears
 one ``Higgs doublet'' as a zero mode of $A_5$
 in a non-SUSY theory.}.

The scalar component of $\Sigma$ can take the 
 vacuum expectation value (VEV) 
 written as 
 ${1\over g R}\sum_a a_a {\lambda_a \over 2}$. 
We can always take VEV as 
\begin{equation}
  \VEV{\Sigma}
    =\frac{1}{2gR}\Ls
    \begin{array}{ccc}
      0 & 0 & a \\
      0 & 0 & 0 \\
      a & 0 & 0 
    \end{array}\Rs
\label{vev-su3}
\end{equation}
 by using the residual $SU(2)\times U(1)$ 
 {\it global} symmetry%
\footnote{
Since this is the D-flat direction, there does not appear
the tree-level quartic couplings in the effective potential. 
We can show that the vacuum exists on the D-flat direction under the 
Scherk-Schwarz SUSY breaking. 
Some low energy contributions might make the vacuum
off the direction, but we assume it is not so large.
Anyway, our scenario shows tan$\beta \simeq 1$. 
}. 
We adopt Scherk-Schwarz (SS) SUSY 
 breaking\cite{SS,SS2,SS3,SS4}. 
The effective potential 
 induced from the gauge sector is given 
 by\cite{HHKY,HY-proof} 
\begin{eqnarray}
V_{\eff}^{gauge}
&=& -2 C \sum_{n=1}^{\infty}{1\over n^5}
    (1-\cos(2\pi n\beta)) [\cos(2 \pi n a)+2\cos(\pi n a)],
\label{26}
\end{eqnarray}
where $C \equiv 3/(64\pi^7R^5)$. 
The $\beta$ parameterizes SS SUSY breaking, 
 for which we take $\beta /R={\mathcal O}(100)$ GeV, since 
 the soft mass is given by 
 ${\mathcal O}(\beta /R)$\cite{HHHK}.

Next, let us calculate 
 the effective potential 
 induced from the quarks and leptons in the bulk.  
As in Ref.\cite{Burdman:2002se}, 
 we introduce 
 ${\bf 3}$, ${\bf 6}$, ${\bf 10}$, and ${\bf 8}$
 representation hypermultiplets 
 of bulk quarks and leptons 
 in order 
 to reproduce the Yukawa 
 interactions of 
 up-, down-, 
 charged lepton-,
 and 
 neutrino-sectors, respectively. 
They all possess $\eta \eta'=+$ 
 in Eqs.(\ref{6}) and (\ref{9}). 
Their contributions to the effective potential 
 are given by 
\begin{eqnarray}
V_{\eff}^{q/l}
&=& 2N_g C\sum_{n=1}^{\infty}{1\over n^5}
         (1-\cos(2\pi n\beta)) \times 
 [3f_u(a)+3f_d(a)+f_e(a)+f_\nu(a)], \\
&& f_u(a)=\cos(\pi n a), \\
&& f_d(a)=\cos(2\pi n a)+\cos(\pi n a), \\
\label{17}
&& f_e(a)=\cos(3\pi n a)+\cos(2\pi n a)+2\cos(\pi n a), \\
&& f_\nu(a)=\cos(2\pi n a)+2\cos(\pi n a), 
\label{18}
\end{eqnarray}
where $N_g$ is the generation number of 
 quarks and leptons in the bulk,  
 and 
 $f_u(a)$, $f_d(a)$, $f_e(a)$, and $f_\nu(a)$,
 are contribution from 
 up-, down-, 
 charged lepton-,
 and 
 neutrino-sectors, 
 respectively. 
The coefficients 3 in front of $f_u(a)$ and $f_d(a)$ 
 denote the color factors.
The contributions from the fundamental, {\bf 3}, 
 the symmetric tensor, {\bf 6}, and the adjoint, {\bf8}, 
 representations are shown in 
 the general formula 
 in Ref.\cite{HY-proof}. 
The remaining contribution from 
 ${\bf 10}$ representation can be 
 calculated by use of 
 the calculation method in Ref.\cite{HY-proof} 
 as follows.

The VEV in Eq.(\ref{vev-su3}) 
 is proportional to one generator 
 of $SU(2)_{13}$ 
 that operates on the $2\times2$ submatrix  
 of $(1,1)$, $(1,3)$, $(3,1)$ and $(3,3)$ 
 components. 
The ${\bf 10}$ is decomposed 
 as ${\bf 4+3+2+1}$ under 
 the base of $SU(2)_{13}$. 
The $U(1)$ charge for the VEV direction, 
 Eq.(\ref{vev-su3}), 
 is given by 
\begin{equation}
\label{charge-ad-su3}
(\underbrace{+3/2, +1/2, -1/2, -3/2}_{\bf 4}, 
 \underbrace{+1, -1, 0}_{\bf 3}, 
 \underbrace{+1/2, -1/2}_{\bf 2}, 
 \underbrace{\ 0\phantom{,} }_{\bf 1}) .
\end{equation}
This means that the eigenvalues of 
 $D_y(A_5)^2$ for a {\bf 10} representation field $B$ are 
\begin{equation}
 2\times {n^2\over R^2}, \;\;\;
 {(n\pm 3a/2)^2\over R^2}, \;\;\;
 {(n\pm a)^2\over R^2}, \;\;\;
 2\times{(n\pm a/2)^2\over R^2}.
\end{equation}
Here eigenfunctions of $B$ are expanded as
 $B\propto\cos{ny\over R},\sin{ny\over R}$, 
 since $(P,P')$ of 
 components of $B$ are either
 $(+,+)$ or $(-,-)$. 
Therefore, the contribution from the 
 {\bf 10} representation to the 
 effective potential is 
\begin{eqnarray}
  &&{i\over2}\int \frac{{\rm d}^4p}{(2\pi)^4}{1\over 2\pi R}
    \sum_{n=-\infty}^\infty
    \Ll   \ln \Ls -p^2+\Ls\frac{n}{R}\Rs^2 \Rs
        + \ln \Ls -p^2+\Ls\frac{n-3a/2}{R}\Rs^2 \Rs \RR \nn\\
  &&\;\;\;\;\;\;\;\;\;\;\;\;\;\;\;\;\;\;\;\;\;\;\;\;\;\;
    \LL + \ln \Ls -p^2+\Ls\frac{n-a}{R}\Rs^2 \Rs  
        +2\ln \Ls -p^2+\Ls\frac{n-a/2}{R}\Rs^2 \Rs
    \Rl . \nn \\
  &&\;\;\;\;\;\;={C\over2} \sum_{n=1}^\infty \;{1\over n^5}\;
     [\cos(3\pi na)+\cos(2\pi na)+2\cos(\pi na)],
\end{eqnarray}
up to $a$-independent terms, 
 for one degree of freedom 
 of fermion. 
This is the derivation of Eq.(\ref{17}).

Then, the total effective potential 
 is given by 
\begin{eqnarray}
V_{\eff}&=&V_{\eff}^{gauge}+V_{\eff}^{q/l} 
= 2C\sum_{n=1}^{\infty}{1\over n^5}
         (1-\cos(2\pi n\beta)) \times \nn \\
&& \;\;\;\;\;\;\;\;\;\;\;
 [N_g\cos(3\pi n a)+(5N_g-1)\cos(2\pi n a)+(10N_g-2)\cos(\pi na)] 
\label{Vgqlsu3}
\end{eqnarray}
Seeing 
 the 1st derivative of 
 $V_{\eff}$, 
 each term of ${\partial V_{\eff}/ \partial a}$ 
 has a factor 
    $\sin(\pi na)$,
 which 
 means that the stationary points exist
 at least at $a=0$ and $a=1$\footnote{ 
The potential has the symmetry
 $V_{\eff}(-a)=V_{\eff}(a)$, 
 so that we should only check the region of $0\leq a \leq 1$.}. 
The difference of the heights 
 between two points is given by 
\begin{eqnarray}
&&V_{\eff}(a=0)-V_{\eff}(a=1)= 
4(11N_g-2)C 
  \sum_{n=1}^{\infty}
 {1 \over (2n-1)^5}
 (1-\cos(2\pi (2n-1)\beta)) .
\label{d11}
\end{eqnarray}
This means that 
 the height of 
 $a=0$ point is higher than 
 that of $a=1$ 
 even when there are one generation 
 quarks and leptons in the bulk. 
The vacuum at $a=1$ point induces 
 the Wilson loop, 
\begin{eqnarray}
W_C&=&
\exp \Ls ig \int_{0}^{2 \pi R}dy \VEV{\Sigma} \Rs \nonumber \\
&=&
\exp \Ls ig 2\pi R{1\over 2g R}
 \Ls\begin{array}{ccc}
  0&0&1 \\ 0&0&0 \\ 1&0&0
 \end{array}\Rs \Rs
=
\left(
\begin{array}{ccc}
-1 & & \\
   &1& \\
   & &-1 
\end{array}
\right),  
\label{WC1su3}
\end{eqnarray}
which suggests $SU(2)_L\times U(1)_Y$ is broken down 
 to $U(1)_\EM\times U(1)$ at the energy scale of  
 ${\mathcal O}(R^{-1})$. 
So this case can not 
 reproduce the correct 
 weak scale VEV, since the 
 compactification scale should be 
 higher than the weak scale. 
Notice here that it does not mean 
 the model in Ref.\cite{Burdman:2002se} 
 is incorrect. 
There 
 the suitable 
 background VEV is introduced in order to make 
 bulk fields localized 
 at $a=0$ and $a=1$   
 and to achieve the fermion mass 
 hierarchy. 
The authors also introduce 
 the wall-bulk mixing
 mass terms, which 
 make unwanted zero modes, 
 such as triplet of $SU(3)_W$ in 
 {\bf 6}, be heavy. 
Our calculations 
 do not include these effects,  
 so that our analyses of vacuum above
 are not for the model 
 in Ref.\cite{Burdman:2002se}. 
In this paper, we take the standing 
 point that the fermion mass hierarchy
 should be reproduced by another mechanism 
 (as that shown in Ref.\cite{HS}, where 
 localization of bulk fields is not needed),  
 and neglect the effects of wall-bulk mixing
 mass terms, which is 
 justified 
 when the compactification scale 
 is higher than 
 these mass terms\footnote{
The effect to the effective potential from wall-bulk mixing 
mass terms can be vanished due to the parity assignment.
We thank N. Okada for pointing 
 out this. 
}. 
In this situation, 
 we can conclude that 
  the vacuum does not realize 
 the suitable electro-weak symmetry 
 breaking. 
So this model should be modified as 
 realizing the suitable vacuum 
 for 
 the 
 realistic gauge-Higgs unification 
 scenario.

We introduce extra fields in the 
 bulk for the suitable 
 electro-weak symmetry breaking. 
We should impose a discrete symmetry 
 in order to avoid unwanted Yukawa
 interactions 
 between ordinary particles and 
 extra fields. 
We consider the following two possibilities; 
\begin{enumerate}
 \item The one loop effective potential chooses 
 $a=0$ vacuum at the high energy 
 (compactification) scale, 
 and the electro-weak symmetry breaking 
 is realized by other effects, such as \cite{ph-gh}, 
 in the low energy.
 \item In the TeV scale 
 compactification, 
 the suitable electro-weak
 symmetry breaking 
 is obtained
 through the 
 one loop effective potential. 
\end{enumerate}
We realize these two situations 
 by introducing extra bulk fields of 
 $N_f^{(\pm)}$ and $N_a^{(\pm)}$ species of 
 hypermultiplets of fundamental and 
 adjoint representations, respectively\cite{HHKY}, 
 where index $(\pm)$ denotes 
 the sign of $\eta \eta'$.  
We should take even number of $N_f^{(\pm)}$ 
 to avoid the gauge anomaly. 
The extra matter contributions for the effective potential 
 are given by\cite{HHKY,HY-proof}
\begin{eqnarray}
V_{\eff}^{extra-m}
&=& 2 C\sum_{n=1}^{\infty}{1\over n^5}
         (1-\cos(2\pi n\beta)) \times 
 \Ll N_a^{(+)}\cos(2\pi na)+N_a^{(-)}
  \cos(\pi n(2a-1))\nonumber \RR\\
&&\LL+  \Ls 2N_a^{(+)}+N_f^{(+)} \Rs \cos(\pi na)+
       \Ls 2N_a^{(-)}+N_f^{(-)}\Rs \cos(\pi n(a-1))\Rl. 
\label{Vm}
\end{eqnarray}
The total effective potential 
 is $V_{\eff}=V_{\eff}^{gauge}+V_{\eff}^{q/l}+V_{\eff}^{extra-m}$. 
The vacuum of this effective potential 
 has the 
\begin{eqnarray}
\label{phase}
&& {\rm (1): \;unbroken \;phase}\; (a=0), \nonumber \\ 
&& {\rm (2): \;broken \;phase \;I}\; (a=1), \\
&& {\rm (3): \;broken \;phase \;I\!I}\; (a\neq0,1),\nonumber
\end{eqnarray}
 in which remaining gauge symmetries 
 are (1): $SU(2)_L\times U(1)_Y$, 
 (2): $U(1)_\EM\times U(1)$, 
 and (3): $U(1)_\EM$, 
 respectively. 
We can see 1st derivatives of $V_\eff^{extra-m}$ 
 at $a=0$ and $a=1$ vanish, 
 thus, that of the total effective potential also dose.
The stability of each stationary point is determined
 by the 2nd derivative of the effective potential
 evaluated at the point 
 as examined in Ref.\cite{Takenaga:2003dp}.
By using the approximation formula 
 for a small (positive) $\xi$, 
\begin{eqnarray}
\label{ap1}
  && \sum^{\infty}_{n=1} 
     {\cos(n\xi) \over n^3}\simeq 
     \zeta(3)+{\xi^2 \over 2}\ln\xi-{3\over4}\xi^2, \\
  && \sum^{\infty}_{n=1} 
     {\cos(n\xi) \over n^3}(-1)^n\simeq 
     -{3\over4}\zeta(3)+{\xi^2 \over 2}\ln2, 
\label{ap2}
\end{eqnarray}
 where $\zeta_R(z)$ is the Riemann's zeta function, 
 the 2nd derivatives are approximated as
\beqn
\left.{\partial^2 V_{\eff} \over \partial a^2}\right|_{a=0}
&=& -2C\pi^2\sum_{n=1}^{\infty}{1\over n^3}
            \Ls 1-\cos(2\pi n\beta)\Rs 
    \Ll 9N_g + 4\Ls 5N_g-1+N_a^{(+)} \Rs + 
     4N_a^{(-)} (-1)^n 
\RR\nonumber \\
&&\hsp{2}\LL +\Ls 10N_g -2 + 2 N_a^{(+)} + N_f^{(+)} \Rs 
             +\Ls 2 N_a^{(-)} + N_f^{(-)}\Rs (-1)^n
\Rl \nn\\
&\simeq&2C\pi^2(2\pi\beta)^2 
  \Ll \Ls 39N_g-6+6N_a^{(+)}+ N_f^{(+)}\Rs 
      \Ls -\half\ln(2\pi \beta)+{3\over4}\Rs
\RR\nn\\
&&\hsp{2}\LL +\Ls -6N_a^{(-)}- N_f^{(-)}\Rs \half\ln2 \Rl,
\label{M0susysu3}\\ 
\left.{\partial^2 V_{\eff} \over \partial a^2}\right|_{a=1}
&=& -2C\pi^2\sum_{n=1}^{\infty}{1\over n^3}
            \Ls 1-\cos(2\pi n\beta)\Rs 
    \Ll 9N_g(-1)^n + 4\Ls 5N_g-1+N_a^{(+)}\Rs + 
     4N_a^{(-)} (-1)^n 
\RR\nonumber \\
&&\hsp{2}\LL +\Ls 10N_g -2 + 2 N_a^{(+)} + N_f^{(+)} \Rs (-1)^n 
             +\Ls 2 N_a^{(-)} + N_f^{(-)}\Rs 
\Rl \nn\\
&\simeq&2C\pi^2(2\pi\beta)^2 
  \Ll \Ls 20N_g-4+4N_a^{(+)}+ 2N_a^{(-)} + N_f^{(-)}\Rs 
      \Ls -\half\ln(2\pi \beta)+{3\over4}\Rs
\RR\nn\\
&&\hsp{2}\LL +\Ls -19N_g+2-2N_a^{(+)}-4N_a^{(-)}- N_f^{(+)}\Rs 
              \half\ln2 \Rl.           
\label{M1susysu3}
\eeqn
The point $a=0\,(a=1)$ is stable 
 when $\left.{\partial^2 V_{\eff} \over 
             \partial a^2}\right|_{a=0} >0$
 $\Ls\left.{\partial^2 V_{\eff} \over 
          \partial a^2}\right|_{a=1} >0\Rs$.  
Whether these points become the 
 true vacuum (the global minimum) or not, 
 more detailed analyses are needed 
 since there is the possibility to find 
 local minimums at other points, in general. 
However, the numerical analyses show 
 the following results, at least, 
 when the numbers of bulk fields 
 are not extremely large. 
\bit
 \item When $\left.{\partial^2 V_{\eff} \over 
                    \partial a^2}\right|_{a=0} >0$
        and $\left.{\partial^2 V_{\eff} \over 
                    \partial a^2}\right|_{a=1} <0$, 
        the stationary point $a=0$ becomes a global 
        minimum, 
        and the unbroken phase 
        is realized. 
 \item When $\left.{\partial^2 V_{\eff} \over 
                    \partial a^2}\right|_{a=0} <0$
        and $\left.{\partial^2 V_{\eff} \over 
                    \partial a^2}\right|_{a=1} >0$, 
        the stationary point $a=1$ becomes a global minimum, 
        and the broken phase I 
        is realized.
 \item When $\left.{\partial^2 V_{\eff} \over 
                    \partial a^2}\right|_{a=0} >0$
        and $\left.{\partial^2 V_{\eff} \over 
                     \partial a^2}\right|_{a=1} >0$, 
        either two vacua are degenerated as a global minimum or 
        one of the points of $a=0$ and $a=1$ becomes a global minimum,
        depending on bulk fields contents. 
\eit
They can be understood 
 by the following discussion. 
Since 
 $n=1$ (of the summation of $n$) 
 and $n=2$ (when contributions of $n=1$ are canceled 
  between $\eta\eta'=+$ part and $\eta\eta'=-$ part.)
 have dominant contributions to 
 the form of the effective potential, 
 the effective potential 
 is approximated as a sum of $\pm\cos(\pi a)$, 
 $\pm\cos(2\pi a)$, $\cos(3\pi a)$ and $\cos(4\pi a)$.
When, for example, $\left.{\partial^2 V_{\eff} \over 
                           \partial a^2}\right|_{a=0}>0$, 
 $-\cos(\pi a)$ and/or $-\cos(2\pi a)$ must dominate 
 the effective potential.
This means that there is no global
 minimum 
 between $a=0$ and $a=1$, when 
 $\left.{\partial^2 V_{\eff} \over 
             \partial a^2}\right|_{a=0}$
 and/or $\left.{\partial^2 V_{\eff} \over 
             \partial a^2}\right|_{a=1}$
 are positive. 
It is justified as long as 
 there are no 
 higher representation fields in the bulk 
 that induce 
 the other terms (\eg\ $-\cos(3\pi a)$).

The remaining case is as follows. 
\bit
 \item When $\left.{\partial^2 V_{\eff} \over 
                    \partial a^2}\right|_{a=0} <0$
        and $\left.{\partial^2 V_{\eff} \over 
                    \partial a^2}\right|_{a=1} <0$, 
        there is a global minimum at $a\neq0,1$, 
        and the broken phase \II\ is realized. 
\eit
It is worth noting 
 that 
 the phase structure 
 is completely determined by 
 the bulk fields contents in this situation.

Above discussion suggests  
 that we can obtain a vanishing VEV, $a=0$, 
 as the global minimum when 
 $\left.{\partial^2 V_{\eff} \over 
         \partial a^2}\right|_{a=0} >0$ and 
 $\left.V_{\eff}\right|_{a=0} <\left.V_{\eff}\right|_{a=1}$.
And, we can obtain a small VEV, $(0<)a\ll1$, 
 when we choose a point near the region 
 corresponding to the unbroken phase 
 in the parameter space.
Namely, when $\left.{\partial^2 V_{\eff} \over 
             \partial a^2}\right|_{a=0} \lesssim0$
 and $\left.{\partial^2 V_{\eff} \over 
          \partial a^2}\right|_{a=1} <0$, 
 a small VEV will be obtained%
\footnote{
In Ref.\cite{HHKY} 
 we have obtained the bulk field content 
 by finding the case 
 that 
 induce large coefficients of $-\cos(\pi na)$ 
 and/or $\cos(\pi n(a-1))$, 
 and 
 small (but non-zero) coefficients of 
 $\cos(2\pi na)$ and $-\cos(2\pi n(a-1/2))$.  
This approach of finding 
 the suitable bulk field content inducing 
 $a$ $(a \ll 1)$ is 
 generalized 
 by 
 the approach of finding 
 the cases of $\left.{\partial^2 V_{\eff} \over 
             \partial a^2}\right|_{a=0} \lesssim0$
 and $\left.{\partial^2 V_{\eff} \over 
          \partial a^2}\right|_{a=1} <0$. 
}.

\begin{figure}
\begin{center}
\includegraphics[width=5cm]{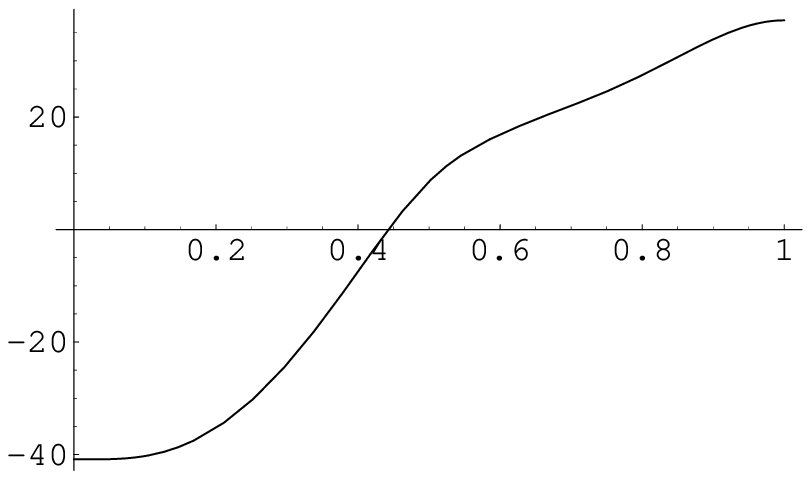}
\hspace{1cm}
\includegraphics[width=5cm]{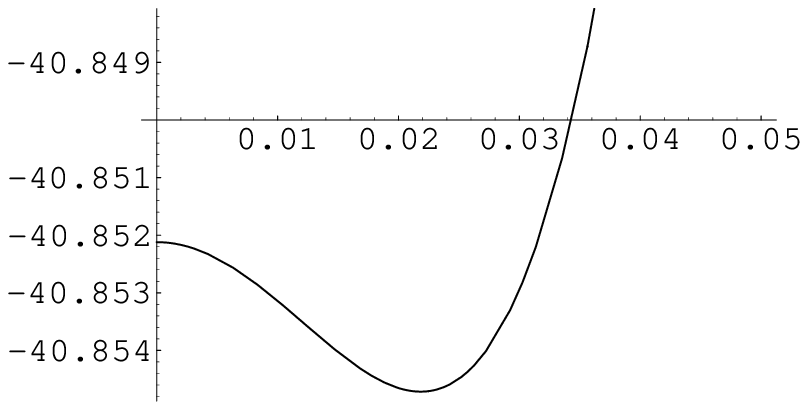}
\caption{The effective potential in the case of 
 $N_a^{(+)}=N_f^{(+)}=0$, $N_a^{(-)}=45$, $N_f^{(-)}=40$
 with $\beta=0.1$ and $N_g=3$.
The unit is $C=3/64\pi^7R^5$. 
The horizontal line shows $0\leq a \leq 1$ and $0\leq a \leq 0.05$. }
\label{fig:SUSYSU3}
\end{center}
\end{figure}

Now let us consider the situation 
 all three generation quarks and leptons 
 exist in the bulk. 
Equation (\ref{M0susysu3}) suggests
 that 
 at least ${\mathcal O}(10)$ 
 numbers 
 of $N_a^{(-)}$ and/or $N_f^{(-)}$ are required
 for realizing vacuum at $a=0$ and $a\ll 1$, 
 since 
 $\Ls-\half\ln(2\pi\beta)+\frac{3}{4}\Rs/\Ls\half\ln2\Rs
  \sim3$ for $\beta=0.1$. 
The former situation, $a=0$,  
 can be realized 
 when the number of extra bulk fields are
 $N_a^{(+)}=0$, $N_f^{(+)}=0$, 
 $N_a^{(-)}=50$, and $N_f^{(-)}=50$, 
 for examples. 
For realizing the latter situation, 
 we assume that 
 the compactification scale
 is of ${\mathcal O}(1)$ 
 TeV\footnote{In this paper we assume 
 that the effect of wall-localized 
 kinetic terms to the effective potential\cite{CA} 
 is negligible. 
We need to check whether 
 the suitable value 
 of $\sin^2 \theta_W$ and the gauge coupling 
 unification are realized 
 by the effects of wall-localized 
 kinetic terms without affecting 
 the effective potential.
For the possibilities, 
 the power law unification\cite{Dienes:1998vh}
 or the accelerated unification\cite{Arkani-Hamed:2001vr} 
 might be useful.} 
 and 
 wall-bulk mass scale 
 is of 
 ${\mathcal O}(100)$ 
 GeV. 
The
 suitable electro-weak symmetry breaking is
 realized 
 when the global minimum exists at $a \ll 1$, 
 and  
 one example of the parameter set for 
 realizing it
 is 
 $N_a^{(+)}=0$, $N_f^{(+)}=0$, 
 $N_a^{(-)}=45$, and $N_f^{(-)}=40$. 
Figure \ref{fig:SUSYSU3} 
 shows the 
 $V_{\eff}$ in the 
 region of $0\leq a \leq 1$ 
 and $0\leq a \leq 0.05$. 
The minimum exists at $a=0.022$, which is around
 the suitable magnitude of the weak scale in the 
 case of TeV scale compactification. 
The kinetic term of the ``Higgs field'' 
 can be reproduced through the effects of 
 wall-localized kinetic terms\cite{Burdman:2002se}. 
Setting $ \left<A_5\right>/{\sqrt{2\pi R}}
 ={a}/{g_4R}\sim246{\rm GeV}$, 
the mass squared of the ``Higgs field'' is given by 
\begin{equation} 
m_{A_5}^2 = (gR)^2 
     \left.{\partial^2 V_{\eff} \over \partial a^2}\right|_{a=0.022}
 = {3g_4^2 \over 32 \pi^4 R^2}
     \left.{\partial^2 (V_{\eff}/(C\pi^2)) \over 
            \partial a^2}\right|_{a=0.022}.
\end{equation}
Numerical analysis shows 
\begin{equation} 
m_{A_5}^2 \sim \left({0.062 g_4 \over R}\right)^2 
   \sim (700 g_4^2 \; {\rm GeV})^2, 
\end{equation}
where the 4D gauge coupling constant 
 $g_4 \equiv g/\sqrt{2\pi R}$ is 
 expected to be of  
 ${\mathcal O}(1)$. 
We take $\beta = 0.1$, since 
 the soft mass is given by 
 $\beta /R$. 
We should notice that 
 the suitable global minimum 
 can be realized by introducing 
 less numbers of extra bulk fields,
 when only the 2nd and 3rd generations, or 
 only the 3rd generation quarks and leptons 
 spread in the bulk.

\section{$SU(6)$ model}

Next, we study the 
 vacuum structure 
 of the $SU(6)$ model,
 in which 
 the Higgs doublets can be identified
 as the zero mode 
 components of $\Sigma$\cite{Hall:2001zb,Burdman:2002se}. 
We take 
\begin{eqnarray}
&&P=\diag(1,1,1,1,-1,-1), 
\label{P}\\
&&P'=\diag(1,-1,-1,-1,-1,-1),
\label{P'}
\end{eqnarray}
in the base of $SU(6)$, 
 which divide $V$ and $\Sigma$ as
\begin{eqnarray}
&&V=\left(
\begin{array}{c|ccc|cc}
(+,+)&(+,-)&(+,-)&(+,-)&(-,-)&(-,-)\\ \cline{1-6}
(+,-)&(+,+)&(+,+)&(+,+)&(-,+)&(-,+)\\ 
(+,-)&(+,+)&(+,+)&(+,+)&(-,+)&(-,+)\\
(+,-)&(+,+)&(+,+)&(+,+)&(-,+)&(-,+)\\ \cline{1-6}
(-,-)&(-,+)&(-,+)&(-,+)&(+,+)&(+,+)\\
(-,-)&(-,+)&(-,+)&(-,+)&(+,+)&(+,+)
\end{array}
\right),\\
&&\Sigma =\left(
\begin{array}{c|ccc|cc}
(-,-)&(-,+)&(-,+)&(-,+)&(+,+)&(+,+)\\ \cline{1-6}
(-,+)&(-,-)&(-,-)&(-,-)&(+,-)&(+,-)\\
(-,+)&(-,-)&(-,-)&(-,-)&(+,-)&(+,-)\\
(-,+)&(-,-)&(-,-)&(-,-)&(+,-)&(+,-)\\ \cline{1-6}
(+,+)&(+,-)&(+,-)&(+,-)&(-,-)&(-,-)\\
(+,+)&(+,-)&(+,-)&(+,-)&(-,-)&(-,-)
\end{array}
\right).
\end{eqnarray}
They suggest that 
 $P$ and $P'$ make $SU(6)$ break to 
 $SU(3)_c\times SU(2)_L\times U(1)_Y \times U(1)$. 
Also, there appears 
 two ``Higgs doublet'' superfields 
 as the zero modes of 
 $\Sigma$.  
The VEV of $\Sigma$ is written as 
\begin{equation}
\label{36}
{1\over 2gR}
    \Ls
    \begin{array}{cccccc}
      0 & 0 & 0 & 0 & 0 & a \\
      0 & 0 & 0 & 0 & 0 & 0 \\
      0 & 0 & 0 & 0 & 0 & 0 \\
      0 & 0 & 0 & 0 & 0 & 0 \\
      0 & 0 & 0 & 0 & 0 & 0 \\
      a & 0 & 0 & 0 & 0 & 0 
    \end{array}\Rs
 \equiv \frac{1}{gR}a\frac{\lambda}{2}
\label{VEVSU6}
\end{equation}
by using the residual $SU(2)_L\times U(1)_Y\times U(1)$ 
 {\it global} symmetry. 
The gauge contribution of the 
 effective potential in the $SU(6)$ model 
 is given by\cite{HHKY,HY-proof} 
\begin{eqnarray}
V_{\eff}^{gauge}
&=& -{2}C\sum_{n=1}^{\infty}{1\over n^5}
    (1-\cos(2\pi n\beta))\times \\
& & [6\cos (\pi n (a-1))+2\cos(\pi
    na)+\cos(2\pi na)] . \nonumber
\end{eqnarray}

As for the quarks and leptons,  
 we introduce 
 ${\bf 20}$, ${\bf 15}$, ${\bf 15}$, and ${\bf 6}$
 representation bulk hypermultiplets in order 
 to reproduce the Yukawa 
 interactions of 
 up-, down-,
 charged lepton-,
 and 
 neutrino-sectors, 
 respectively\cite{Burdman:2002se}. 
The up- and down-sector fields, 
 ${\bf 20}$ and ${\bf 15}$, 
 possess $\eta \eta'=-$, 
 while 
 the charged lepton- and neutrino-sector fields, 
 ${\bf 15}$ and ${\bf 6}$, 
 possess $\eta \eta'=+$\cite{Burdman:2002se}. 
Their contributions to the effective potential 
 are given by 
\begin{eqnarray}
V_{\eff}^{q/l}
&=& 2N_g C\sum_{n=1}^{\infty}{1\over n^5}
         (1-\cos(2\pi n\beta)) \times 
 [g_u(a)+g_d(a)+g_e(a)+g_\nu(a)], \\
\label{39}
&& g_u(a)=3\cos(\pi n a)+3\cos(\pi n (a-1)), \\
&& g_d(a)=3\cos(\pi n a)+2\cos(\pi n (a-1)), \\
&& g_e(a)=2\cos(\pi n a)+3\cos(\pi n (a-1)), \\
&& g_\nu(a)=\cos(\pi n a), 
\label{Vqlsu6}
\end{eqnarray}
where $N_g$ is the generation number, and 
 $g_u(a)$, $g_d(a)$, $g_e(a)$, and $g_\nu(a)$,
 are contribution from 
 up-, down-,
 charged lepton-,
 and 
 neutrino-sectors, 
 respectively. 
The contributions from the fundamental, {\bf 6}, 
 and anti-symmetric tensor, {\bf 15}, representations 
 are shown in 
 the general formula in Ref.\cite{HY-proof}. 
The remaining contributions from {\bf 20} can 
 be also calculated by using 
 the same calculation method in 
 Ref.\cite{HY-proof}. 
As suggested there, the product of parities 
 (Eq.(\ref{P}) and Eq.(\ref{P'})), 
\bequ
 PP'=\diag(1,-1,-1,-1,1,1),
\eequ 
by which the gauge symmetry is reduced as 
 $SU(6)\rightarrow SU(3)_c\times SU(3)_L\times U(1)$, 
 plays an important role. 
In this case, the generator $\lambda$ in Eq.(\ref{VEVSU6}) 
 is one of the generators of $SU(3)_L$.
The ${\bf 20}$ is decomposed as 
\bequ
 {\bf35} \rightarrow 
         ({\bf3} ,{\bf{\bar3}},+)+({\bf1},{\bf1},+)
        +({\bf{\bar3}},{\bf{3}},-)+({\bf1},{\bf1},-) 
\eequ
in terms of ($SU(3)_c$, $SU(3)_L$, $PP'$). 
This means that the eigenvalues of 
 $D_y(A_5)^2$ for a {\bf20} representation field $B$ are
\begin{equation}
 8\times {n^2\over R^2}, \;\;\;
 3\times{(n\pm a/2)^2\over R^2}, \;\;\;
 3\times{(n\pm a/2-1/2)^2\over R^2}, 
\end{equation}
where eigenfunctions of $B$ are expanded as 
 $B\propto\cos{ny\over R}, \sin{ny\over R}$ 
 $(\cos{(n+1/2)y\over R}, \sin{(n+1/2)y\over R})$ 
 for $(P,P')=(+,+)$ or $(-,-)$ 
 ($(+,-)$ or $(-,+)$). 
Therefore, the contribution from the 
 {\bf 20} representation to the effective potential 
 is 
\begin{eqnarray}
  &&{i\over2}\int \frac{{\rm d}^4p}{(2\pi)^4}{1\over 2\pi R}
    \sum_{n=-\infty}^\infty
    \Ll 4 \ln \Ls -p^2+\Ls\frac{n}{R}\Rs^2 \Rs
       +3 \ln \Ls -p^2+\Ls\frac{n-a/2}{R}\Rs^2 \Rs \RR \nn\\
  &&\;\;\;\;\;\;\;\;\;\;\;\;\;\;\;\;\;\;\;\;\;\;\;\;\;\;
    \LL +3\ln \Ls -p^2+\Ls\frac{n-a/2-1/2}{R}\Rs^2 \Rs  
    \Rl . \nn \\
  &&\;\;\;\;\;\;={1\over2}C \sum_{n=1}^\infty \;{1\over n^5}\;
     [3\cos(\pi na)+3\cos(\pi n(a-1))],
\end{eqnarray}
up to $a$-independent terms, for one degree of freedom 
 of the fermion.
This is just the Eq.(\ref{39}).

Now, we have the total effective potential, 
\begin{eqnarray}
V_{\eff}&=&V_{\eff}^{gauge}+V_{\eff}^{q/l} 
= 2C\sum_{n=1}^{\infty}{1\over n^5}
         (1-\cos(2\pi n\beta)) \times \nn \\
&& \;\;\;\;\;\;\;\;\;\;\;
 [-\cos(2\pi na)+(9N_g-2)\cos(\pi n a)+
 (8N_g-6)\cos(\pi n(a-1))] .
\label{Vgqlsu6}
\end{eqnarray}
As in the previous section, 
 the 1st derivative, 
 ${\partial V_{\eff}/ \partial a}$ 
 has a factor 
    $\sin(\pi na)$,
 which 
 means that the stationary points exist
 at least at $a=0$ and $a=1$. 
The difference of the heights 
 between two points is given by 
\begin{eqnarray}
&&V_{\eff}(a=0)-V_{\eff}(a=1)= 
 4(N_g+4)C 
  \sum_{n=1}^{\infty}
 {1 \over (2n-1)^5}
 (1-\cos(2\pi (2n-1)\beta)) .
\label{d1}
\end{eqnarray}
This means that 
 the height of 
 $a=0$ is higher than 
 that of $a=1$ even when 
 $N_g=0$. 
The vacuum $a=1$ induces 
 the Wilson loop, 
\begin{eqnarray}
W_C&=&
\exp (ig \int_{0}^{2 \pi R}dy{1\over g R} a 
 {\lambda\over 2} ) \nonumber \\
&=&
\exp (ig {1\over g R}{\lambda\over 2}2\pi R)
=
\left(
\begin{array}{cccccc}
-1 & & & & & \\
   &1& & & & \\
   & &1& & & \\
   & & &1& & \\
   & & & &1& \\
   & & & & &-1 
\end{array}
\right),  
\label{WC1su6}
\end{eqnarray}
which suggests $SU(2)_L\times U(1)_Y$ is broken to 
 $U(1)_\EM\times U(1)$. 
This case can not reproduce the correct 
 weak scale VEV as in 
 the previous 
 section.

Then, let us 
 introduce extra fields in the 
 bulk for the suitable 
 electro-weak symmetry breaking. 
We introduce the extra bulk fields of 
 $N_f^{(\pm)}$ and $N_a^{(\pm)}$ species of 
 hypermultiplets of fundamental and 
 adjoint representations, respectively. 
The extra matter contributions for 
 the effective potential 
 are given by\cite{HHKY,HY-proof}
\begin{eqnarray}
V_{\eff}^{extra-m}
&=& 2C\sum_{n=1}^{\infty}{1\over n^5}(1-\cos(2\pi n\beta))\Ll
    N_a^{(+)}\cos(2\pi na)+
    N_a^{(-)}\cos(\pi n(2a-1)) \RR\nonumber \\ 
&& +\Ls 2N_a^{(+)}+6N_a^{(-)}+N_f^{(+)}\Rs
     \cos(\pi na) \nonumber \\ 
&& \LL+\Ls 6N_a^{(+)}+2N_a^{(-)}+N_f^{(-)}\Rs\cos(\pi n(a-1))\Rl.
\end{eqnarray}
There are three phases as Eq.(\ref{phase}). 
As in the previous section, 
 the phase structure is obtained by 
 the bulk field content, which is 
 calculated by the 2nd derivatives at $a=0$ and $a=1$, 
\beqn
\left.{\partial^2 V_{\eff} \over \partial a^2}\right|_{a=0}
&\simeq&2C\pi^2(2\pi\beta)^2 
  \Ll \Ls 9N_g-6+6N_a^{(+)}+6N_a^{(-)}+ N_f^{(+)} \Rs 
       \Ls -\half\ln(2\pi \beta)+{3\over4}\Rs
\RR\nn\\
&&\hsp{2}\LL  + \Ls -8N_g+6-6N_a^{(+)}-6N_a^{(-)}- N_f^{(-)} \Rs 
       \half\ln2 \Rl,
\label{M0susysu6}\\ 
\left.{\partial^2 V_{\eff} \over \partial a^2}\right|_{a=1}
&\simeq&2C\pi^2(2\pi\beta)^2 
  \Ll \Ls 8N_g-10+10N_a^{(+)}+2N_a^{(-)}+ N_f^{(-)} \Rs 
       \Ls -\half\ln(2\pi \beta)+{3\over4}\Rs
\RR\nn\\
&&\hsp{2}\LL  + \Ls -9N_g+2-2N_a^{(+)}-10N_a^{(-)}- N_f^{(+)} \Rs 
       \half\ln2 \Rl,
\label{M1susysu6}
\eeqn
Here 
 we use Eqs.(\ref{ap1}) and (\ref{ap2}). 
As mentioned in the previous section, 
 $a=0$ point 
 becomes 
 the global minimum when 
 $\left.{\partial^2 V_{\eff} \over 
         \partial a^2}\right|_{a=0} >0$ and 
 $\left.V_{\eff}\right|_{a=0} <\left.V_{\eff}\right|_{a=1}$.
On the other hand, 
 a small VEV, $(0<)a\ll1$ is realized 
 when $\left.{\partial^2 V_{\eff} \over 
             \partial a^2}\right|_{a=0} \lesssim0$
 and $\left.{\partial^2 V_{\eff} \over 
          \partial a^2}\right|_{a=1} <0$.

\begin{figure}
\begin{center}
\includegraphics[width=5cm]{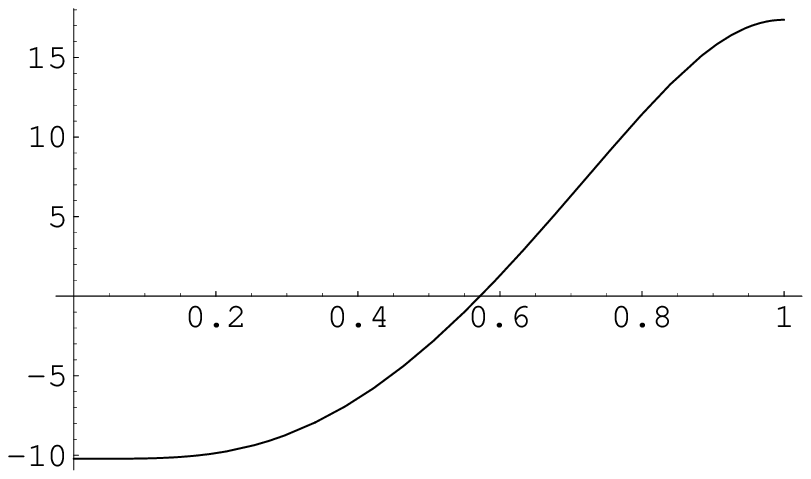}
\hspace{1cm}
\includegraphics[width=5cm]{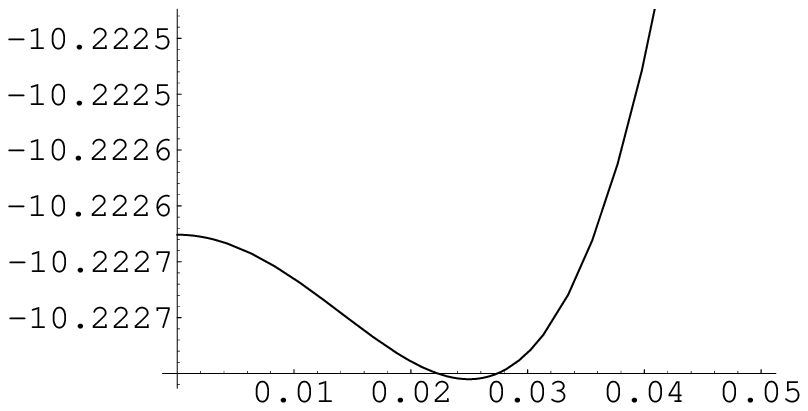}
\caption{The effective potential in the case of 
 $N_a^{(+)}=N_f^{(+)}=0$, $N_a^{(-)}=0$, $N_f^{(-)}=42$
 with $\beta=0.1$ and $N_g=3$.
The unit is $C=3/64\pi^7R^5$. 
The horizontal line shows $0\leq a \leq 1$ and $0\leq a \leq 0.05$. }
\label{fig:SUSYSU6}
\end{center}
\end{figure}

Let us consider the situation 
 all three generation quarks and leptons 
 exist in the bulk. 
For both cases 
 at least ${\mathcal O}(10)$ 
 numbers of $N_f^{(-)}$ are 
 required from Eq.(\ref{M0susysu6}).
The vacuum, $a=0$, is 
 realized, for instance, when 
 the extra bulk fields are 
 introduced as 
 $N_a^{(+)}=0$, $N_f^{(+)}=0$, 
 $N_a^{(-)}=0$, and $N_f^{(-)}=50$.  
While 
 in the TeV scale compactification, 
 the global minimum $a \ll 1$ is possible  
 when, for examples, 
 $N_a^{(+)}=0$, $N_f^{(+)}=0$, 
 $N_a^{(-)}=0$, $N_f^{(-)}=42$. 
Figure \ref{fig:SUSYSU6} 
 shows the 
 $V_{\eff}$ in the 
 region of $0\leq a \leq 1$ 
 and $0\leq a \leq 0.05$. 
The minimum exists at $a=0.025$, which is around
 the suitable magnitude of the weak scale in the 
 TeV scale compactification. 
The mass squared of the ``Higgs field'' is given by 
\begin{equation} 
m_{A_5}^2 \sim \left({0.012 g_4 \over R}\right)^2 
   \sim (120 g_4^2 \; {\rm GeV})^2, 
\end{equation}
where $g_4 = {\mathcal O}(1)$ and 
 $\beta = 0.1$. 

As in the previous section, 
 the suitable vacuum 
 can be obtained by 
 less numbers of extra bulk fields
 when only the 2nd and 3rd generations, or 
 only the 3rd generation quarks and leptons 
 exist in the bulk.
As for 
 the extra residual $U(1)$ gauge symmetry, 
 we assume 
 it 
 is broken by 
 an extra elementally Higgs field.

\section{Summary and discussion}

We have studied the possibility of the dynamical 
 symmetry breaking in the 
 gauge-Higgs unification 
 in the 5D ${\mathcal N}=1$ 
 SUSY 
 theory 
 compactified on an orbifold, $S^1/Z_2$. 
We have considered $SU(3)_c \times SU(3)_W$ and 
 $SU(6)$ models, where 
 the gauge symmetries are reduced 
 to $SU(3)_c \times SU(2)_L \times U(1)_Y$ and 
 $SU(3)_c \times SU(2)_L \times U(1)_Y \times U(1)$,
 respectively, 
 through 
 the orbifolding boundary conditions. 
Our setup is quarks and leptons are 
 bulk fields, so that 
 Yukawa interactions 
 can be derived from 
 the 5D gauge interactions. 
We have calculated the one loop effective 
 potential
 of ``Higgs doublets''
 and analyzed the 
 vacuum structure 
 of the models. 
We found that the effects of bulk quarks and leptons
 destabilize the suitable electro-weak symmetry
 breaking vacuum.
We showed that the introduction of 
 suitable numbers of 
 extra bulk fields 
 possessing the suitable representations  
 makes two appropriate scenarios be possible. 
One is the situation that the 
 one loop effective potential chooses 
 symmetric vacuum at the 
 compactification scale, 
 and the electro-weak symmetry breaking 
 is occurred by other effects in 
 the low energy. 
The other is the situation that the 
 one loop effective potential chooses 
  the suitable electro-weak 
 vacuum 
 in a few TeV compactification, where  
 the masses of ``Higgs doublets'' are 
 $\mathcal{O}(100)$ GeV. 
In this case we can obtain the suitable electro-weak symmetry breaking 
 with vanishing tree-level quartic couplings. 
It is because the one loop effective potentials have the form 
 of cos-function, which is the characteristic feature 
 of the Wilson line phase. 
And also the introduction of the bulk fields with 
 $\eta\eta'=-$ is crucial for inducing 
 the electro-weak energy scale\cite{HHKY},
 which is smaller than the compactification scale. 
The large number of bulk fields, $N_{a,f}^{(-)}={\mathcal O}(10)$, 
 makes the "Higgs" mass rather large, 
 although there are no tree-level quartic couplings 
 in this model.

\vskip 1.5cm

\acknowledgments

We would like to thank Y. Hosotani, Y. Kawamura, and
 K. Takenaga  
 for helpful discussions.  
This work was supported in part by  Scientific Grants from 
 the Ministry of Education and Science, 
 Grant No.\ 14039207, 
 Grant No.\ 14046208, \ Grant No.\ 14740164 (N.H.), and 
 by a Grant-in-Aid for 
 the 21st Century COE ``Center for Diversity 
 and Universality in Physics'' (T.Y.).

\vskip 1cm

\def\jnl#1#2#3#4{{#1}{\bf #2} (#4) #3}

\def\Zphys{{\em Z.\ Phys.} }
\def\jssc{{\em J.\ Solid State Chem.\ }}
\def\jpsJ{{\em J.\ Phys.\ Soc.\ Japan }}
\def\ptps{{\em Prog.\ Theoret.\ Phys.\ Suppl.\ }}
\def\PTP{{\em Prog.\ Theoret.\ Phys.\  }}

\def\JMP{{\em J. Math.\ Phys.} }
\def\NPB{{\em Nucl.\ Phys.} B}
\def\NP{{\em Nucl.\ Phys.} }
\def\PLB{{\em Phys.\ Lett.} B}
\def\PL{{\em Phys.\ Lett.} }
\def\PRL{\em Phys.\ Rev.\ Lett. }
\def\PRB{{\em Phys.\ Rev.} B}
\def\PRD{{\em Phys.\ Rev.} D}
\def\PRe{{\em Phys.\ Rep.} }
\def\AP{{\em Ann.\ Phys.\ (N.Y.)} }
\def\RMP{{\
em Rev.\ Mod.\ Phys.} }
\def\ZPC{{\em Z.\ Phys.} C}
\def\SCI{\em Science}
\def\CMP{\em Comm.\ Math.\ Phys. }
\def\MPLA{{\em Mod.\ Phys.\ Lett.} A}
\def\IJMPA{{\em Int.\ J.\ Mod.\ Phys.} A}
\def\IJMPB{{\em Int.\ J.\ Mod.\ Phys.} B}
\def\EPJC{{\em Eur.\ Phys.\ J.} C}
\def\PR{{\em Phys.\ Rev.} }
\def\JHEP{{\em JHEP} }
\def\cmp{{\em Com.\ Math.\ Phys.}}
\def\JPA{{\em J.\  Phys.} A}
\def\CQG{\em Class.\ Quant.\ Grav. }
\def\ATMP{{\em Adv.\ Theoret.\ Math.\ Phys.} }
\def\ibid{{\em ibid.} }

\leftline{\bf References}

\renewenvironment{thebibliography}[1]
         {\begin{list}{[$\,$\arabic{enumi}$\,$]}  
         {\usecounter{enumi}\setlength{\parsep}{0pt}
          \setlength{\itemsep}{0pt}  \renewcommand{\baselinestretch}{1.2}
          \settowidth
         {\labelwidth}{#1 ~ ~}\sloppy}}{\end{list}}


\begin{thebibliography}{99}
\small
\baselineskip=14pt







\bibitem{Manton:1979kb}
N.~S.~Manton,
Nucl.\ Phys.\ B {\bf 158}, (1979), 141;\\
D.~B.~Fairlie,
J.\ Phys.\ G {\bf 5}, (1979), L55;
Phys.\ Lett.\ B {\bf 82}, (1979), 97.


\bibitem{YH}
Y.~Hosotani, Phys. Lett. {\bf B126} ({1983}), {309}; 
Ann. of Phys. {\bf 190} ({1989}), {233}; 
Phys. Lett. {\bf B129} ({1984}), {193}; 
Phys. Rev. {\bf D29} ({1984}), {731}. 



\bibitem{Krasnikov:dt}
N.~V.~Krasnikov,
Phys.\ Lett.\ B {\bf 273}, (1991), 246;\\
H.~Hatanaka, T.~Inami and C.~S.~Lim,
Mod.\ Phys.\ Lett.\ A {\bf 13}, (1998), 2601;\\
G.~R.~Dvali, S.~Randjbar-Daemi and R.~Tabbash,
Phys.\ Rev.\ D {\bf 65}, (2002), 064021;\\
N.~Arkani-Hamed, A.~G.~Cohen and H.~Georgi,
Phys.\ Lett.\ B {\bf 513}, (2001), 232;\\
I.~Antoniadis, K.~Benakli and M.~Quiros,
New J.\ Phys.\  {\bf 3}, (2001), 20.


\bibitem{Lim}
M.~Kubo, C.~S.~Lim and H.~Yamashita,
Mod.\ Phys.\ Lett.\ A {\bf 17} (2002), 2249.  


\bibitem{Hall:2001zb}
L.~J.~Hall, Y.~Nomura and D.~R.~Smith,
Nucl.\ Phys.\ B {\bf 639}, 307 (2002).


\bibitem{Burdman:2002se}
G.~Burdman and Y.~Nomura,
Nucl.\ Phys.\ B {\bf 656} (2003), 3. 


\bibitem{HS}
N.~Haba and Y.~Shimizu, 
Phys.\ Rev.\ D {\bf 67} (2003), 095001. 


\bibitem{CA}
C. A. Scrucca, M. Serone and L. Silvestrini,
Nucl.\ Phys.\ B {\bf 669} (2003) 128.


\bibitem{mimura1}
I.~Gogoladze, Y.~Mimura and S.~Nandi,
Phys.\ Lett.\ B {\bf 560} (2003), 204; 
Phys.\ Lett.\ B {\bf 562} (2003), 307;\\
I.~Gogoladze, Y.~Mimura, S.~Nandi and K.~Tobe,
Phys.\ Lett.\ B {\bf 575} (2003), 66.  

 
\bibitem{ghu3}
C.~Csaki, C.~Grojean, H.~Murayama, L.~Pilo and J.~Terning,
hep-ph/0305237.


\bibitem{ph-gh}
K.~Choi, N.~Haba, K.~S.~Jeong, K.~i.~Okumura, 
 Y.~Shimizu and M.~Yamaguchi,
JHEP {\bf 0402} (2004), 037.



\bibitem{HHKY}
N. Haba, Y. Hosotani, Y. Kawamura, and T. Yamashita, 
hep-ph/0401183. 



\bibitem{HY-proof}
%
%
N. Haba 
and T. Yamashita, 
JHEP {\bf 0402} (2004), 059.


\bibitem{HHHK}
N. Haba, M. Harada, Y. Hosotani, and Y. Kawamura, 
Nucl.\ Phys.\ {\bf 657} (2003), 169. 


\bibitem{HHK}
N. Haba, Y. Hosotani, and Y. Kawamura, 
Prog. Theor. Phys. {\bf 111} (2004), 265. 



\bibitem{6DLag}
N.~Arkani-Hamed, T.~Gregoire and J.~Wacker,
JHEP {\bf 0203} (2002), 055.


\bibitem{Kaplan:2001ga}
D.~E.~Kaplan and T.~M.~Tait,
JHEP {\bf 0111} (2001) 051.













\bibitem{SS}
J.~Scherk and J.~H.~Schwarz,
Phys.\ Lett.\ B {\bf 82} (1979), 60; Nucl.\ phys.\ B {\bf 153} (1979),
61.

\bibitem{SS2}
P.\ Fayet,
\jnl{\PLB}{159}{121}{1985}; \jnl{\NPB}{263}{649}{1986}.


\bibitem{SS3}
I. Antoniadis, 
{\it Phys. Lett.} {\bf B246} (1990), 377;\\ 
I. Antoniadis, C. Munoz, and M. Quiros, 
{\it Nucl. Phys.} {\bf B397} (1993), 515;\\ 
A.\ Pomarol and M.\ Quiros, \jnl{\PLB}{438}{255}{1998};\\
I.~Antoniadis, S.~Dimopoulos, A.~Pomarol and M.~Quiros,
{\it Nucl.\ Phys.}\ B {\bf 544}, 503 (1999);\\ 
A.~Delgado, A.~Pomarol and M.~Quiros,
{\it Phys.\ Rev.}\ D {\bf 60}, 095008 (1999). 


\bibitem{SS4}   
G.V.\ Gersdorff,   M.\ Quiros and A.\ Riotto, \jnl{\NPB}{634}{90}{2002};\\
G.V.\ Gersdorff and M.\ Quiros, \jnl{\PRD}{65}{064016}{2002}.


\bibitem{Takenaga:2003dp}
K.~Takenaga,
Phys.\ Lett.\ B {\bf 570} (2003), 244. 



\bibitem{Dienes:1998vh}
K.~R.~Dienes, E.~Dudas and T.~Gherghetta,
Phys.\ Lett.\ B {\bf 436} (1998), 55;\\
K.~R.~Dienes, E.~Dudas and T.~Gherghetta,
Nucl.\ Phys.\ B {\bf 537} (1999), 47.  
  
\bibitem{Arkani-Hamed:2001vr}
N.~Arkani-Hamed, A.~G.~Cohen and H.~Georgi,
hep-th/0108089.

\end{thebibliography}
\end{document}